
\magnification=1200  
\def\newline{\hfill\penalty -10000}  
\def\title #1{\centerline{\rm #1}}
\def\author #1; #2;{\line{} \centerline{#1}\smallskip\centerline{#2}}
\def\abstract #1{\line{} \centerline{ABSTRACT} \line{} #1}
\def\heading #1{\line{}\smallskip \goodbreak \centerline{#1} \line{}}
\newcount\refno \refno=1
\def\refjour #1#2#3#4#5{\noindent    \hangindent=1pc \hangafter=1
 \the\refno.~#1, #2 ${\bf #3}$, #4 (#5). \global\advance\refno by 1\par}
\def\refbookp #1#2#3#4#5{\noindent \hangindent=1pc \hangafter=1
 \the\refno.~#1, #2 (#3, #4), p.~#5.    \global\advance\refno by 1\par}
\def\refbook #1#2#3#4{\noindent      \hangindent=1pc \hangafter=1
 \the\refno.~#1, #2 (#3, #4).           \global\advance\refno by 1\par}
\newcount\equatno \equatno=1
\def\adveqn{(\the\equatno) \global\advance\equatno by 1}

\def\up#1{\leavevmode \raise 0.2ex\hbox{#1}}

\newcount\figno
\figno=0
\def\figure{\global\advance\figno by 1 Figure~\the\figno.~}
%

%
\vsize=8.75truein
\hsize=5.75truein
\hoffset=0.5truein
%
\baselineskip=0.166666truein
%
\parindent=25pt
%
\parskip=0pt
%
\nopagenumbers

%
%
\line{}

\title{
ON THE NATURE OF NONTHERMAL RADIATION FROM
}
\title{
COSMOLOGICAL $\gamma$-RAY BURSTERS
}

\author
Vladimir V. Usov \parindent=0pt ;
Dept.~of Physics, Weizmann Institute, Rehovot 76100, Israel ;

\abstract{
Relativistic electron-positron winds with strong magnetic fields
are considered as a source of radiation for cosmological
$\gamma$-ray bursters. Such a wind is generated by a
millisecond pulsar with a very strong magnetic field.
An electron-positron plasma near the pulsar is optically
thick and in quasi-thermodynamic equilibrium. It is shown that
the main part of radiation from the pulsar wind is nonthermal
and generates in the following way. Kinetic energy which is released
in the process of deceleration of the neutron star rotation
transforms mainly to the magnetic field energy. The magnetic field is
frozen in the outflowing plasma if the distance to the pulsar is
smaller than $\sim 10^{13}$ cm. This field transfers the energy from
the pulsar environment to the region outside
the $\gamma$-ray photosphere of the
electron-positron wind. At a distance of more than $\sim 10^{13}$
cm the magnetohydrodynamic approximation for the pulsar wind is
broken, and intense electromagnetic waves are generated. The
frequency of these waves is equal to the frequency of the pulsar
rotation. Outflowing particles are accelerated in the field of
intense electromagnetic waves to Lorentz factors of the order of
$10^6$ and generate nonthermal synchro-Compton radiation.
The typical energy of nonthermal photons is $\sim 1$ MeV.
A high-energy tail of the $\gamma$-ray spectrum may be
up to $\sim 10^4$ MeV. Baryonic matter is ejected occasionally
from the pulsar magnetosphere. The baryonic matter ejection and
subsequent suppression of the $\gamma$-ray emission may be responsible
for the time structure of $\gamma$-ray bursts.
} 

\heading{INTRODUCTION}
The BATSE experiment on board the {\it Compton Gamma Ray Observatory}
found that the $\gamma$-ray burst sources are distributed isotropically
in the sky, and it sees the edge of the distribution$^1$.
These two facts can be naturally explained if $\gamma$-ray
bursters are at cosmological distances$^{2-8}$.
Besides, there are other indications of a cosmological
origin for $\gamma$-ray bursts (see ref. 7
and references therein).
\par A generic problem with a cosmological model is that the huge
initial energy density implies a very large optical depth to
electron-positron pairs, thermalization of the electron-positron
plasma, and a blackbody spectrum with small modifications$^{9,10}$.
This is a clear conflict with the observed spectra
of $\gamma$-ray bursts, which are well fitted either as power
laws or as broken power laws$^{11,12}$.
It was suggested$^{13,14}$ that a very strong magnetic field
may be in the electron-positron plasma which flows away from the
$\gamma$-ray burster. Here it is shown that outflowing particles
may be accelerated outside the $\gamma$-ray photospheres of
relativistic electron-positron winds with strong magnetic fields.
These particles may be responsible for the nonthermal radiation
of $\gamma$-ray bursts.

\heading{PARTICLE ACCELERATION AND $\gamma$-RAY EMISSION}

Below, I assume that millisecond pulsars with extremely
strong magnetic fields, $B_{_S} \simeq$ a few
$\times 10^{15}$ G, are a cosmological source of $\gamma$-ray
bursts$^{13}$. In such a model the neutron star rotation is
a source of energy for $\gamma$-ray bursts. The rotation of
the magnetic neutron star decelerates because of the electromagnetic
torque. The rate of kinetic energy loss for a supposedly nearly
orthogonal magnetic dipole is$^{15}$
$$- {d E_{\rm kin} \over dt}\simeq L_{\rm md}= {2\over 3}
{B^2_{_S}R^6\Omega ^4 \over c^3}\simeq 2 \times 10 ^{51}
\left({B_{_S}\over 3\times 10^{15}\, {\rm G}}\right)^2
\left({\Omega \over 10^4 \, {\rm s}^{-1}}\right)^4 \,\,{\rm
erg}\,{\rm s}^{-1}\, , \eqno (1)$$
where $E_{\rm kin}$ is the rotational kinetic energy, $R\simeq 10^6$ cm
is the radius of the neutron star and $\Omega$ is the angular velocity.
\par Like ordinary  known pulsars,
electron-positron pairs are created in the magnetosphere of a
millisecond pulsar with a very strong magnetic field.
The electron-positron plasma and radiation are
in quasi-thermodynamical equilibrium in the environment
of such a pulsar$^{13}$.
\par The electron-positron plasma with strong magnetic fields
flows away from the pulsar at relativistic speeds.
During outflow, the electron-positron plasma accelerates and its
density decreases. At a distance $r_{\rm ph}$ from the pulsar where the
optical depth for the main part of photons is $\tau _{\rm ph}
\sim 1$, the radiation propagates freely. If we don't take into
account the magnetic field, the radius of the $\gamma$-ray photosphere
for a spherical optically thick
electron-positron wind is$^9$
$$r_{\rm ph} \simeq \left({L_p\over 4\pi\sigma T^4_{_0}\Gamma^2_{\rm ph}
}\right)^{1/2}\,,\eqno (2)$$
where $T_{_0}$ is the temperature of electron-positron plasma at
$r\simeq r_{\rm ph}$ in the co-moving frame, $\sigma = 5.67\times
10^{-5}$ erg cm$^{-2}$ K$^{-4}$ s$^{-1}$ is the
Stefan-Boltzmann constant and $\Gamma _{\rm ph}$
is the mean Lorentz factor of plasma particles at $r\simeq r_{\rm ph}$.
\par Since $\Gamma _{\rm ph} \simeq r_{\rm ph}/r_{lc}$ and $T_{_0}\simeq
2\times 10^8$ K (ref. 3), we have
$$r_{\rm ph}\simeq \left({L_pr_{lc}^2 \over 4\pi\sigma T^4_{0}}
\right)^{1/4}\simeq 3.5\times 10^8 \alpha ^{1/4}
\left({B_{_S}\over 3\times 10^{15}
\, {\rm G}}\right)^{1/2}\left({\Omega\over 10^4 \,{\rm s}^{-1}}
\right)^{1/2}\,\,{\rm cm}\,, \eqno (3)$$
$$\Gamma _{\rm ph} \simeq 10^2\alpha ^{1/4}
\left({B_{_S}\over 3\times 10^{15} \,{\rm G}}\right)^{1/2}
\left({\Omega \over 10^4 \,{\rm s}^{-1}}\right)^{3/2}\,,\eqno (4)$$
where $r_{lc} = c/\Omega$ is the radius of the pulsar light cylinder.
\par At $r < r_{\rm ph}$ the optical depth increases sharply
with decreasing $r$. Therefore, any energy which is inherited
by particles and radiation at a distance to the pulsar a few times
smaller than $r_{\rm ph}$
will be thermalized before it is radiated at $r\simeq r_{\rm ph}$.
\par The luminosity of the $\gamma$-ray photosphere is $\alpha L_
{\rm md}\sim (0.01-0.1)L_{\rm md}$. The temperature which
corresponds to the blackbody-like radiation from the $\gamma$-ray
photosphere is $\sim 2\Gamma _{\rm ph}T_{_0}\sim 10^{10}$ K,
and the typical energy of $\gamma$-rays is $\sim 1$ MeV.
\par Kinetic energy which is released in the process
of deceleration of the neutron star rotation transforms mainly to the
magnetic field energy but not to the energy of particles$^{16,17}$.
The pulsar luminosity in magnetic fields is $(1 - \alpha )L_{\rm md}
\simeq L_{\rm md}$. The strength of the magnetic field which
is generated outside the pulsar light cylinder
because of the neutron star rotation is
$$B\simeq B_{_S}\left({R\over r_{lc}}\right)^3\left({r_{lc}\over r}
\right)\simeq 3.3\times 10^{14}{R\over r}\left({B_{_S}\over 3\times
10^{15}\,{\rm G}}\right)\left({\Omega\over 10^4\,{\rm s}^{-1}}\right)
^2\,\,{\rm G}\,.\eqno (5)$$
The magnetic field is frozen in the outflowing electron-positron
plasma if the distance to the pulsar is not too large (see below).
This field can transfer the energy from the pulsar environment to
the region outside the $\gamma$-ray photosphere, $r > r_{\rm ph}$,
without its thermalization.
\par The magnetic field does not change qualitatively the motion of
relativistic electron-positron wind inside the $\gamma$-ray photosphere.
At $r\simeq r_{\rm ph}$, the density of electrons and positrons
drops sharply, and the particle acceleration because of the
magnetic field may be essential.
\par Acceleration of particles in the pulsar wind at $r > r_{lc}$
is characterized by the following dimensionless parameter$^{18}$
$$\eta ={\Omega ^2\Phi ^2\over 4\pi fc^3}\,, \eqno (6)$$
where
$$f = \rho v_rr^2,\,\,\,\,\,\,\,\,\,\Phi = r^2B_r\,,
\,\,\,\,\,\,\,\,\,\,\rho = n_{_\pm}m\,,\eqno (7)$$
$n_{_\pm}$ is the laboratory frame number density,
$m$ is the mass of electron, $v_r$ is the radial velocity and $B_r$ is
the radial component of the magnetic field.
\par Continuity of the
magnetic flux gives $\Phi$ = constant. At the pulsar
light cylinder, $r = r_{lc}$, we have $B_r\simeq B_{_S}(R/r_{lc})^3$
and $\Phi \simeq B_{_S}R^2(\Omega R/c)$. Using this value of $\Phi$
and taking into account that $v_r\simeq c$ for relativistic flow,
from equations (1), (6) and (7) we obtain
$$\eta \simeq {L_{\rm md}\over mc^2\dot N_{_\pm}}\,,\eqno (8)$$
where $\dot
N_{_\pm}$ is the flux of electrons and positrons from the pulsar.
\par For the electron-positron wind the flux of particles at $r >
r_{\rm ph}$ is$^9$
$$\dot
N_{_\pm}\simeq {4\pi c r_{\rm ph}\Gamma ^2_{\rm ph}\over \sigma _{_T}}
\simeq 2\times 10^{48}\alpha ^{3/4}
\left({B_{_S}\over 3\times 10^{15}\,{\rm G}}
\right)^{3/2}\left({\Omega \over 10^4\,{\rm s}^{-1}}\right)^{7/2}
\,\,{\rm s}^{-1}\,,\eqno (9)$$
where $\sigma _{_T} = 6.65\times 10^{-25}$ cm$^2$ is the Thomson cross
section.
\par Equations (1), (8) and (9) yields:
$$\eta \simeq 10^9\alpha ^{-3/4}
\left({B_{_S}\over 3\times 10^{15}\, {\rm G}}
\right)^{1/2}\left({\Omega \over 10^4\,{\rm s}^{-1}}\right)^{1/2}\,.
\eqno (10)$$
\par The density of electron-positron plasma  at $r\simeq r_{\rm ph}$
$$n_{_\pm} = {\dot N_{_\pm}\over 4\pi c r^2_{\rm ph}}\simeq 4\times
10^{19}\alpha ^{1/4}
\left({B_{_S}\over 3\times 10^{15}\,{\rm G}}\right)^{1/2}
\left({\Omega \over 10^4\,{\rm s}^{-1}}\right)
^{5/2}\,\,{\rm cm}^{-3} \eqno (11)$$
is essentially higher than the critical value$^{18}$
$$n_{cr} = {\Omega B\over 4\pi ce}\simeq 4\times 10^{16}{R\over r}
\left({B_{_S}\over 3\times 10^{15}\,{\rm G}}\right)\left({\Omega
\over 10^4\,{\rm s}^{-1}}\right)^3 \,\,{\rm cm}^{-3}\,. \eqno (12)$$
Therefore, the magnetic field is frozen in the plasma, and
the magnetohydrodynamic (MHD) approximation can be used to describe the
wind motion$^{18}$. In this approximation, particles may be
accelerated to Lorentz factors of the order of $\eta^{1/3}$ (ref. 18),
which is an order of magnitude more than $\Gamma _{\rm ph}$.
However, the $\eta ^{1/3}$ estimate of $\Gamma$ is valid only if
either the interaction between particles and photons is negligible
or the mass density of radiation, $\rho _{\gamma} =aT^4/c^2$, is smaller
than the mass density of particles, $\rho _{_\pm} =n_{_\pm}m$ (here
$a = 7.56\times 10^{-15}$ erg cm$^{-3}$ K$^{-4}$ is radiation density
constant). Near the $\gamma$-ray photosphere the density of
radiation is very high, and the interaction between particles
and radiation is strong. This interaction results in the increase
of the mass density of accelerated matter. Substituting $\rho _\gamma$
for $\rho$ in equation (6), we have the following estimate
for the Lorentz factor of accelerated particles near the $\gamma$-ray
photosphere: $\Gamma _m \simeq ({\eta\rho _{_\pm}/\rho _\gamma })^{1/3}$.
For $B_{_S}\simeq 3\times 10^{15}$ G, $\Omega \simeq 10^4$ s$^{-1}$,
$T=T_{_0}$ and $\alpha \simeq 0.01-0.1$,
we have $\Gamma _m\sim \Gamma _{\rm ph}$. Hence, there is no essential
acceleration of particles because of the magnetic field near
the $\gamma$-ray photosphere in the MHD approximation.
Therefore, the region of the pulsar wind near the $\gamma$-ray
photosphere is not promising for a generation of strong nonthermal
radiation.
\par With the distance from the pulsar the density of particles
decreases in proportion to $r^{-2}$. The critical density decreases
somewhat slower (see equation (12)). At the distance
$$r_{\rm nth}\simeq 1.3\times 10^{14}\alpha ^{3/4}
\left({B_{_S}\over 3\times 10^{15}
\,{\rm G}}\right)^{1/2}\left({\Omega \over 10^4\,{\rm s}^{-1}}\right)
^{1/2}\,\,{\rm cm} \eqno (13)$$
the plasma density, $n_{_\pm}$, is equal to the critical one, $n_{cr}$.
At $r > r_{\rm nth}$ the MHD approximation is broken for the pulsar wind
with the magnetic
field which alternates in polarity on the scalelength of
$\sim \pi (c/\Omega ) \sim 10^7$ cm (refs 18,19),
and intense electromagnetic waves with the frequency of $\Omega$ can
propagate outside$^{19,20}$.
In this case the process of particle acceleration
changes qualitatively, namely: particles can be accelerated in this
wind zone to Lorentz factors of the order of $\eta ^{2/3} \sim 10^6$
(refs 21-23)
in contrast to the $\eta ^{1/3}$ estimate given by the MHD theory.
Particles are accelerated on the scalelength of $\sim r_{\rm nth}$
and generate synchro-Compton radiation. The radiative damping length for
intense electromagnetic waves with the wavelength $\lambda = 2\pi
(c/\Omega)$ is$^{23}$
$$l\simeq {6\over \pi ^4}{\Omega ^3 \over c^3r^3_{_0}n^2_{_\pm}}
\left({n_{cr}\over n_{_\pm}}\right)\,\,\,\,{\rm wavelengths}\,,
\eqno (14)$$
where $r_{_0} = e^2/mc^2 = 2.8\times 10^{-13}$ cm is classical electron
radius.
\par For a millisecond pulsar with a very strong magnetic field,
$B_{_S}\simeq 3\times 10^{15}$ G, $\Omega \simeq 10^4$ s$^{-1}$ and
$\alpha \simeq 0.1$, from equation (14) we have $l\simeq 10^2\lambda
\simeq2\times 10^9$ cm $\ll r_{\rm nth}$ at $r\simeq
2 r_{\rm nth}$. Hence, at a distance of the order of $r_{\rm nth} \sim
10^{13}$ cm the energy of intense electromagnetic waves is transformed
to both the energy of accelerated particles and the energy of nonthermal
synchro-Compton radiation. The luminosity in accelerated particles
at $r > r_{\rm nth}$ is $mc^2\Gamma \dot N_{_\pm}\simeq mc^2\eta ^{2/3}
\dot N_{_\pm}\simeq \eta ^{-1/3}L_{\rm md}\sim 10^{-3}L_{\rm md} \ll
L_{\rm md}$ (see equation (8)).
Therefore, low-frequency intense electromagnetic waves
have to be reradiated into nonthermal high-frequency emission
with the luminosity up to $(1 - \alpha ) L_{\rm md}\simeq L_{\rm md}$.
The typical energy of nonthermal photons, $\epsilon _\gamma \simeq
(\pi /4)\hbar\Omega\eta ^2 \sim 1$ MeV (ref. 23),
is suitable to be identified with the energy of breaks
which are observed in the spectra of $\gamma$-ray bursts$^{11,12}$.
A long high-energy tail of the $\gamma$-ray spectrum may be up to
$\sim \hbar (eB/mc)\eta ^{4/3} \sim 10^4$ MeV. A detailed study
of the $\gamma$-ray spectrum of electron-positron winds with
strong magnetic fields is beyond the framework of this paper and will
be addressed elsewhere.
\par Until now, I have considered relativistic winds which
consist of a pure electron-positron plasma. However, the luminosity
of a very young neutron star is highly super-Eddington, and the
ordinary baryonic matter has to flow away from the neutron star surface.
A characteristic mass-loss rate is $\sim 0.005 M_\odot$ s$^{-1}$
(refs 25,26).
Such a large mass-loss rate almost completely obscures any
prompt electromagnetic display and certainly rules out the production by
any model of $\gamma$-ray bursts situated at cosmological distances.
Extremely strong magnetic fields in the pulsar magnetosphere can
prevent the gas outflow from the neutron star surface threaded by
closed magnetic field lines if the surface temperature
is smaller than the value of $\sim (B_{lc}^2/8\pi a)^{1/4}\simeq 1.5
\times 10^{10}(B_{lc}/10^{14}\,{\rm G})^{1/2}$ K at which the density
of radiation energy, $aT^4$, is equal to the density of magnetic
field energy, $B_{lc}^2/8\pi$, at the pulsar light cylinder.
For $B_{_S} \simeq 3\times 10^{15}$ G, $\Omega \simeq 10^4$ s$^{-1}$
and $B_{lc} \simeq B_{_S}(R/r_{lc})^3\simeq 10^{14}$ G, this upper limit
is a few times higher than the surface temperature of a neutron star
which is formed from an accreting white dwarf$^{25}$.
As to the polar caps near the magnetic poles,
the gas outflow from the neutron star surface in these regions may be
suppressed by the ram pressure of backward flux of particles (Usov, in
preparation) which is predicted by all modern models of
pulsars$^{16,17}$.
\par It is worth noting that
baryonic matter may be ejected occasionally from the neutron
star magnetosphere because of some kind of plasma instabilities
(Usov, in preparation), and the $\gamma$-ray emission may be suppressed
for some time$^{9,24}$.
This process may be responsible for the time structure of $\gamma$-ray
bursts. It is expected that the flux variations in nonthermal
$\gamma$-rays is as short as
$$\tau _{_0} \simeq {r_{\rm nth}\over 2c\Gamma ^2}\simeq {r_{\rm nth}
\over 2c\eta ^{2/3}}\,, \eqno (15)$$
where $\Gamma$, the Lorentz factor of the outflowing plasma particles,
is $\sim \eta ^{1/3}$ at $r_{\rm ph}\ll r < r_{\rm nth}$.
For a pulsar with $B_{_S} \simeq 3\times 10^{15}$ G,
$\Omega \simeq 10^4$ s$^{-1}$ and $\alpha \simeq 0.1$, from equations
(10), (13)  and (15) we have
$\tau _{_0}\simeq 10^{-4}$ s.
\par
The expansion energy of baryonic matter can be reconverted into
nonthermal radiation when it interacts with an external medium$^{27-29}$.
Another component of nonthermal radiation which may be
observed in the burst spectra is annihilation lines$^{30}$.
Positrons which are responsible for these lines can be produced
by burst photons interacting with a medium surrounding a $\gamma$-ray
burster.

\heading{CONCLUSIONS AND DISCUSSION}

I have considered in this paper the radiation from relativistic
electron-positron winds with strong magnetic fields. Such a wind
is generated by a millisecond pulsar with a very strong magnetic field,
and may be responsible for the radiation of cosmological $\gamma$-ray
bursters. Electron-positron plasma near the pulsar is optically thick
and in quasi-thermodynamic equilibrium. It is shown that the
radiation of the pulsar wind consists of two
components. One of them is thermal radiation with a blackbody-like
spectrum. This radiation is emitted by the $\gamma$-ray photosphere
of outflowing electron-positron plasma at the distance of the
order of $10^8$ cm from the pulsar. The other component is
nonthermal synchro-Compton radiation of very high energy
particles which are accelerated at the distance of $\sim 10^{13}$ cm
where the MHD approximation for the pulsar wind is broken and
intense electromagnetic waves with the frequency of $\Omega$
can propagate. For this nonthermal radiation, the characteristic time
of the $\gamma$-ray flux variations is as short as $\sim 10^{-4}$ s.
Besides, an additional component of
nonthermal radiation can be generated because of the
interaction between baryonic matter which is ejected from the
neutron star magnetosphere and an external medium.
For this kind of nonthermal radiation the characteristic
time of flux variations cannot be essentially smaller than one
second$^{28}$.
\par The following correlation between two mentioned
components of nonthermal radiation may be observed for
long bursts. So, if fast variable synchro-Compton radiation from
the electron-positron wind is suppressed by the baryonic matter
ejection, then in a few seconds or more, a slow variable nonthermal
radiation of Rees and M\'esz\'aros (ref. 27)
can appear in the burst spectrum.
\par Most of the results in this paper are applicable for any compact
fast-rotating object for which the rotation decelerates because of the
electromagnetic torque and the rate of kinetic energy loss is high
enough to explain the luminosities of cosmological $\gamma$-ray bursters.
One of such an object may be differentially rotating disks
which are formed by the merger of binaries consisting of either
two neutron stars or a black hole and a neutron star. The strength of
the magnetic field near the postmerger object may be as high as
$\sim 10^{16}-10^{17}$ G (ref. 10).

\heading{ACKNOWLEDGMENTS}

I thank M. Milgrom for helpful conversations and M. Rees for many
helpful suggestions that improved the final manuscript.

\heading{REFERENCES}

\refjour{C.A. Meegan et al.}{Ap.~J.}{355}{143}{1992}
\refjour{V.V. Usov and G.V. Chibisov}{Soviet Astr.}{19}{155}{1975}
\refjour{B. Paczy\'nski}{Ap.~J.}{308}{L43}{1986}
\refjour{B. Paczy\'nski}{Acta Astr.}{41}{257}{1991}
\refjour{T. Piran}{Ap.~J.}{389}{L45}{1992}
\refjour{C.D. Dermer}{Phys. Rev. Lett.}{68}{1799}{1992}
\refjour{W.A.D.T. Wickramasinghe et al}{Ap.~J.}{411}{L55}{1993}
\refjour{P. Tamblyn and F. Melia}{Ap.~J.}{417}{L21}{1993}
\refjour{B. Paczy\'nski}{Ap.~J.}{363}{218}{1990}
\refjour{R. Narayan, B. Paczy\'nski and T. Piran}{Ap.~J.}{395}{L83}{1992}
\refjour{B.E. Schaefer et al.}{Ap.~J.}{393}{L51}{1992}
\refjour{D. Band et al.}{Ap.~J.}{413}{281}{1993}
\refjour{V.V. Usov}{Nature}{357}{472}{1992}
\refjour{C. Thompson and R.C. Dancan}{Ap.~J.}{408}{194}{1993}
\refjour{J.P. Ostriker and J.E. Gunn}{Ap.~J.}{157}{1395}{1969}
\refjour{M.A. Ruderman and P.G. Sutherland}{Ap.~J.}{196}{51}{1975}
\refbookp{J. Arons}{in Proc. Workshop "{\it Plasma Astrophysics}"}
{Varenna, Italy}{1981}{273}
\refjour{F.C. Michel}{Ap.~J.}{158}{727}{1969}
\refjour{V.V. Usov}{Astrophys. Space Sci.}{32}{375}{1975}
\refjour{E. Asseo, F.C. Kennel and R. Pellat}{Astr. Astrophys}
{44}{31}{1975}
\refjour{J.E. Gunn and J.P. Ostriker}{Ap.~J.}{165}{523}{1971}
\refjour{F.C. Michel}{Ap.~J.}{284}{384}{1984}
\refjour{E. Asseo, F.C. Kennel and R. Pellat}{Astr. Astrophys}
{65}{401}{1978}
\refjour{A. Shemi and T. Piran}{Ap.~J.}{365}{L55}{1990}
\refjour{S.E. Woosley and E. Baron}{Ap.~J.}{391}{228}{1992}
\refjour{A. Levinson and D. Eichler}{Preprint}{}{}{1992}
\refjour{M.J. Rees and P. M\'esz\'aros}{MNRAS}{258}{41P}{1992}
\refjour{P. M\'esz\'aros and M.J. Rees}{Ap.~J.}{405}{278}{1993}
\refjour{J. Katz}{Ap.~J.}{}{}{in press}
\refjour{H.S. Fencl, R.N. Boyd and D.H. Hartmann}{Ap.~J.}{407}
{L21}{1993}

\bye